# IP over P2P: Enabling Self-configuring Virtual IP Networks for Grid Computing

Arijit Ganguly, Abhishek Agrawal, P. Oscar Boykin, Renato Figueiredo
Advanced Computing and Information Systems Laboratory
University of Florida
Gainesville, Florida - 32601
email: {aganguly,aagrawal,boykin,renato}@acis.ufl.edu

*Abstract*—**Peer-to-peer (P2P) networks have mostly focused on task oriented networking, where networks are constructed for single applications, i.e. file-sharing, DNS caching, etc. In this work, we introduce IPOP, a system for creating virtual IP networks on top of a P2P overlay. IPOP enables seamless access to Grid resources spanning multiple domains by aggregating them into a virtual IP network that is completely isolated from the physical network. The virtual IP network provided by IPOP supports deployment of existing IP-based protocols over a robust, self-configuring P2P overlay. We present implementation details as well as experimental measurement results taken from LAN, WAN, and Planet-Lab tests.**

## I. INTRODUCTION

Through the use of virtual machines, the native computational environment for an application can be instantiated on-demand on any physical resource [43][30][46][48][35]. This flexibility helps overcome the heterogeneity of Grid [23] computing environments by breaking software dependences between hosts and users, and facilitates controlled and secure sharing of resources by leveraging the additional isolation layer enforced by a virtual machine monitor. However, the execution environment of a distributed computing application also entails the network over which it interacts. It is key that bi-directional TCP/IP connectivity be provided in such distributed execution environments to support a wide spectrum of applications. However, the increasing use of Network Address Translation (NAT) and IP firewalls creates a situation that some nodes on the network can create outgoing connections, but cannot receive incoming connections. This breaks the original model of each node in the Internet being a peer, and is recognized as a hindrance to programming and deploying Grid computing systems [21][49][50]. Protocols for NAT/Firewall traversals [45] exist, but require applications to be re-linked with the new protocol libraries.

Network virtualization techniques for Grid computing have been shown to provide applications their native network environments, despite the idiosyncrasies of the real physical network [10][32]. All complications relating to NAT/Firewall traversals can be handled by the virtualization layer, enabling Grid applications to leverage from a wealth of IP-based software typically available in local-area environments. The core technique employed by these and other virtual networking approaches (e.g. VPNs [2]) is tunneling of virtual network traffic over an IP-overlay. To be deployed in a Grid context, it is desirable that such an overlay is *scalable and fault-tolerant* and that it *requires minimal administrative control*.

In this paper, we present IPOP - a network virtualization technique based on IP tunneling over peer-to-peer (P2P) networks. P2P networks can be made self-configuring, allow user mobility, are scalable, and provide extremely robust service, motivating the choice of P2P routing as the basis for IPOP. Through IPOP, resources spanning multiple domains can be aggregated into a virtual IP network providing bidirectional connectivity. Our protocols support seamless, self-configured addition of nodes to a virtual IP network. Our work might be classified as a P2P protocol for VPN (virtual private networks). The IPOP virtual IP address space is routable *within* the P2P overlay, however it is decoupled from the address space of the physical Internet infrastructure — IP packets are payloads that tunnel through the P2P overlay, as depicted in Figure 1.

Current network virtualization techniques for Grid computing [13][32] require an administrator to setup overlay routing tables. Hence, the process of adding, configuring and managing clients and servers that route traffic within the overlay is difficult to scale. Although topology adaptation is possible using techniques proposed in [10], adaptive routes are coordinated by a centralized server. Both VNET and VIOLIN can provide a robust overlay through redundancy. However, the effort required to preserve robustness would increase every time a new node is added and the network grows in size. In contrast, the use of P2P routing to overlay virtual IP traffic differentiates IPOP from existing network virtualization solutions with respect to the following issues:

- **Scalability**: Network management in a P2P-based routing overlay scales to large numbers because

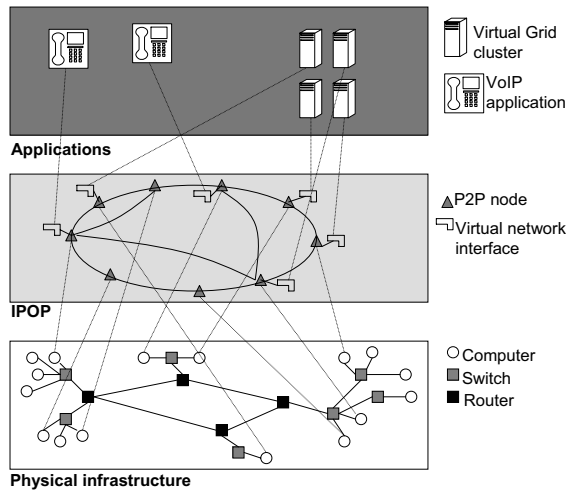

Fig. 1. IPOP: Virtualizing IP over a P2P overlay. The IPOP layer sits between applications (e.g. Grid clusters of physical and/or virtual machines, voice-over-IP) and physical computing nodes interconnected by existing IP networking infrastructures.

routing information is naturally *self-configured*, *decentralized*, and dynamically *adaptive* in response to nodes joining and leaving. Adding a new resource to the virtual IP network requires minimal effort, which is independent of current size of the network. Performance scaling leverages from the fact that each node contributes bandwidth and processing power so that the resources of a system grow as nodes join.

- **Resiliency**: P2P networks are robust even in the face of high failure rates [11][44]. An IP-over-P2P overlay benefits from the synergy of fault-tolerant techniques applied at different levels. The IPOP overlay dynamically adapts routing of IP packets as nodes fail or leave the network; even if packets are dropped by such nodes, IP and other protocols above it in the network stack have been designed to cope with such transient failures. The use of P2P overlays to provide robust routing has also been described in [12].
- **Accessibility**: Existing network virtualization techniques can leverage mechanisms described in [45][25] to cross NAT/Firewalls. These approaches require setting up globally reachable STUN or STUNT servers that aid building the necessary NAT state by carefully crafted exchange of packets. With P2P networks, each overlay node can provide this functionality for detection of NATs and their subsequent traversal. This approach is decentralized and introduces no dedicated servers.

In summary, this paper describes the IPOP architecture and evaluates the performance of a prototype implementation that uses Brunet [16] for P2P routing and "tap" virtual network interfaces. Experiments in local- and wide-area networks are used to quantify the latency and bandwidth characteristics of an IPOP link. Results obtained using "ping" and "ttcp" benchmarks show that the latency overhead (on an implementation that has not been optimized for IP tunneling) is in the range of 6 to 10 ms.

Experiments also show that IPOP overlays can be self-configured, even when nodes are behind firewalls/private networks, and that unmodified applications can use the IPOP overlay as if they were running on a local area network. This paper presents a case study of running an unmodified parallel application (Light Scattering Spectroscopy analysis [41]) that uses TCP/IP services including SSH, message-passing interface (MPI) libraries, and Network File System (NFS) mounted volumes across three firewalled domains in a wide-area network. The experiment shows that the virtual network is successfully self-configured in a decentralized manner and efficiently support the execution of this application.

This paper is organized as follows. Section II discusses related work in grid and peer-to-peer computing. Section III describes the IPOP architecture and its current prototype. Section IV reports on quantitative analyses of the prototype's performance for benchmark applications in local- and wide-area environments. Section V provides ideas for improving IPOP routing performance, followed by Section VI that summarizes the paper.

## II. RELATED WORK

The IPOP system builds on previous research in P2P overlay networks, but our main interest is in applying it to virtual IP networks for Grid computing. In [24], Foster et al. make a case for the convergence of P2P and Grid technologies. In this paper we make the case for one important instance where such convergence is beneficial. This section discusses prior works in both areas.

### A. Peer-to-peer

There is an existing body of research on various ways in which P2P systems can be applied to existing IP systems. In [18] Cox et. al. have proposed to build a Distributed DNS using DHash, a peer-to-peer distributed table built on top of Chord [52]. In [28] Hsieh et. al. have argued that TCP is inappropriate for effective data transport over peer-to-peer networks and have suggested the need of transport layer support for multipoint-to-point connections in P2P networks. While we do not disagree that TCP is sub-optimal for distributed file downloads, we do believe that there are important applications where pairs of nodes need to communicate using existing TCP-based implementations, e.g. message-passing libraries and high-throughput task farming in Grid computing.

This paper is also related to the growing body of knowledge in search and routing in P2P networks. A comprehensive survey of search methods can be found in Risson et. al. [31]. Advances in routing techniques that strive to minimize number of hops and account for network performance in selecting hops can be leveraged by IPOP. In [26][27] Gupta et. al. have attempted to build a P2P system which can route lookup queries in just one hop by maintaining complete routing tables at each node. Harvey et. al. [29] have developed SkipNet, a scalable overlay network providing controlled data placement and guaranteed routing locality. Freedman et. al. [38] proposed Coral, which creates self-organizing clusters of nodes that fetch information from each other to avoid communicating with more distant or heavily-loaded servers. Gummadi et. al. [33] have shown that of all the routing geometries, the ring geometry allows greatest flexibility and achieves best resilience and proximity performance.

Zhou et. al. have developed P6P [56], [55], an implementation of IPv6 on a P2P overlay. The Teredo protocol [9] developed by Microsoft tunnels IPv6 packets over IPv4 UDP packets to enable nodes behind NATs to be addressed with IPv6 connectivity. Our approach differs from these works, as our current focus is IPv4 to enable existing grid applications to run unmodified. Few existing applications support IPv6.

The use of P2P based overlay to support legacy applications has also been described in context of i3 ([51][34]). The goal is to support interoperability with new i3 applications that support multicast, anycast and mobility. In contrast, our motivation is to provide seamless access to Grid resources spanning different network domains by aggregating them into a virtual IP network that is completely *isolated* from the physical network.

### B. Grid computing

Many representative efforts on Grid computing have focused on aggregating resources to support high-throughput computing, but at the expense of requiring applications to be designed from scratch [20][37][17][3], constrained to use specialized remote I/O libraries [39] or Grid APIs [22]. IPOP is different as it is directed towards providing a virtual networking platform that enables unmodified applications to run on top of the overlay.

Related projects (VIOLIN [32], VNET [13][10]) have also recognized the utility of network overlays in Grid environments. The performance results reported for these systems show that the virtualization overhead is tolerable in wide-area networks, and have thus motivated the IPOP approach. The difference in IPOP is that nodes joining/leaving the overlay are handled in a completely *decentralized* fashion. In contrast, in VNET and VIOLIN, it is the responsibility of a centralized manager to perform tasks such as setting up of network links and node addresses.

### C. Brunet P2P Overlay

In this work, we make use of the Brunet P2P overlay network [16]. While the techniques will be applicable to many P2P overlays, our experiments are conducted using Brunet. The Brunet system can use either TCP or UDP as its underlying transport. In Section IV we compare the performance of our IP-over-P2P system using both the TCP and UDP modes of Brunet.

Making use of an existing P2P overlay allows us to avoid dealing directly with many of the difficult issues of network virtualization. Specifically, the Brunet library manages the connections, negotiates the firewalls and NAT devices, and guarantees that the network is routable.

## III. IPOP ARCHITECTURE

The IPOP architecture has two key components (Figure 2): a virtualized network interface for capturing and injecting IP packets into the virtual network, and a P2P routing substrate that encapsulates, tunnels and routes packets within the overlay. We will first describe how IP packets are captured from and injected to the host, and then how we route IP packets on the Brunet P2P network.

### A. Capturing IP from the Host

Our IPOP prototype captures packets from a virtual device attached to the source host, tunnels them over the application-level Brunet P2P overlay, and then injects the packet back into a virtual device on the destination host. IPOP uses the *tap* device, which appears as a virtual network interface inside the host, and is available for several Unix platforms [8] and Windows [6]. The prototype currently works for Linux, where *tap* is available for read and write as a character device.

Through *tap*, Ethernet frames injected by the the kernel on the virtual interface can be read by a user-level application, and vice versa. IPOP runs as a C# program over the Brunet library. It reads and writes Ethernet frames from the *tap* device, and uses Brunet to assemble IP packets inside P2P packets and to route them to their destination. While IPOP sees Ethernet frames, it only routes IP packets; non-IP traffic, notably ARP traffic, is contained within the host. A host running IPOP is a P2P node, and can act as data source, receiver and router all at the same time.

Since we are dealing with Ethernet level frames at the IPOP hosts, we may come across non-IP based frames such as ARP and RARP. Our implementation currently is capable of handling ARPs by creating a static ARP entry inside the host for a non-existent "gateway", which routes for all hosts in the virtual address space. The non-existent gateway has a unique IP address in the virtual address space, and its Ethernet address is configured to not conflict with that of the local *tap* device. As a result, it is possible for the host to only send out the IP- based Ethernet frames and contain ARP requests locally. When an IP packet is received at a P2P end-point, an Ethernet packet is built from it with source as the Ethernet address corresponding to the ARP entry for the gateway inside the host, and the destination to be the Ethernet address of the *tap* device on the host.

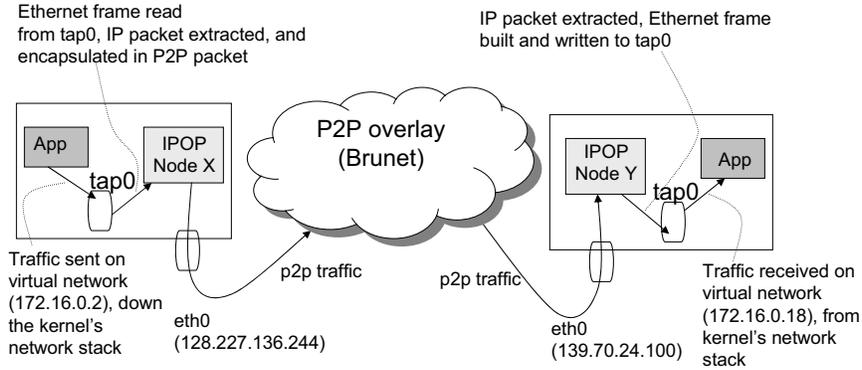

Fig. 2. IPOP architecture overview. Existing IP applications (App) connect to IPOP through a virtual network interface ("tap"). A P2P node (Brunet) extracts IP packets from the virtual interface of a sender (node X, left), and routes them within the overlay to a destination node, where the IP packet is injected into another virtual network interface (node Y, right).

The Mono C# runtime environment used in the current prototype has minimal support for reading and writing character devices such as *tap*. We have thus built a C-based library of low-level functions to open, read, write and close a *tap* device, and use C#'s "PInvoke" feature to invoke them. Our choice of *tap* (layer-2 device) over *tun* (point-to-point device that works at layer-3), is motivated by its extensibility to virtual machines, which we describe in Section III-C.

### B. Routing IP on the P2P Overlay

When a host needs to be added to the IPOP virtual network, the host administrator is required to set up a *tap* device in the host, and bring it up with an IP address that is unique in the virtual IP address space [1]. IPOP runs on the host as a P2P node whose address is the 160 bit SHA-1 hash of the IP address of the *tap* device. Figure 2 shows the traffic flow between two hosts on the virtual network. It shows how applications running on two different hosts communicate over the IPOP virtual network. The application running on Host A generates IP-based traffic which the kernel transmits through the *tap0* interface. The Ethernet frame is captured, an IP packet extracted from it, and then encapsulated inside a P2P packet (Figure 3). The P2P packet, in turn, is sent to the P2P node whose address is the 160 bit SHA-1 hash of the destination IP address. Once the P2P packet is received at the destination node, the IP packet is extracted, an Ethernet frame is built from it, and the frame is written to the *tap0* interface on the host.

IPOP has been designed with the goal of providing open access among a set of trusted nodes spanning multiple sites through network virtualization. When end resources are physical machines, providing an open access through IPOP requires the host firewall rules to be relaxed only for the traffic coming from the virtual interface. No changes are required to the site firewall rules. However, when end resources are virtual machines as described in the next section, we do not even require changes to the host firewall rules.

### C. IPOP extensions for virtual machines

A driving application for IPOP in the context of Grid computing is to interconnect O/S virtual machines (e.g. VMware [53], Xen [14], UML [19]) with the IP overlay, thereby virtualizing key Grid resources [43]. A machine (physical or virtual) running IPOP must have a way of connecting to other P2P nodes, without necessarily having a public IP address. We have implemented and tested our prototype with VMware's bridged and NAT network interfaces. In both cases IPOP runs within the VM. For UML virtual machines, we run IPOP outside the VM "guest" — i.e. in the physical machine's O/S which hosts the VM. In this case, the VM's virtual Ethernet card is configured with an IP address in the virtual address space, and routes and static ARP entries are set as described earlier. The virtual Ethernet card of the UML guest attaches itself to a *tap* device on the host, and the UML kernel uses tap reads and writes to transmit Ethernet frames. A *tap* device can be opened by at most one process, and hence IPOP cannot directly read and write Ethernet frames from the same *tap* device. The *tap* device is therefore bridged [4] to another *tap* device on the host, from which IPOP can read and write Ethernet frames.

---
[1]The MAC address of the interface, however, does not need to be unique.

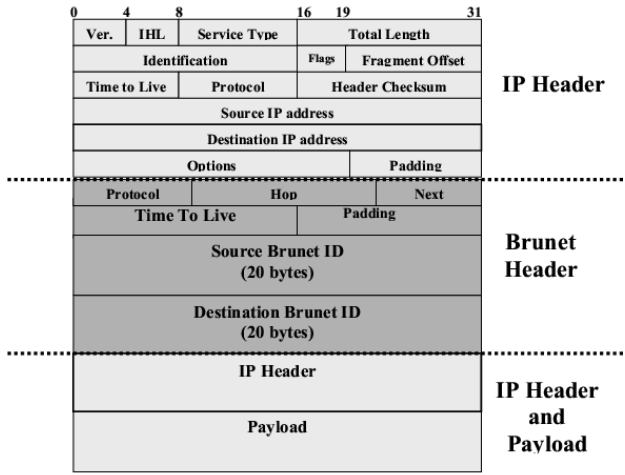

Fig. 3. Structure of an IPOP encapsulated IP packet. The outer IP header is used to transport the Brunet packet to each node in the P2P overlay. The Encapsulated packet is unwrapped when it reaches its final destination.

The advantage of running IPOP inside the VM guest is that no administrative privileges are required on the physical host. However, with UML we still require setting up tap and bridge on the physical host which requires root access. In [32], purely user-level implementation of UML networking switch has been proposed that does not require tap and bridge setup on the physical host; a similar implementation can also be conceived in the context of IPOP.

### D. Crossing firewalls and NATs

One of the many benefits we derive from overlaying IP on a P2P network is the overlay we use, Brunet[16], already deals with connecting nodes which are behind firewalls and NAT devices. In this section we briefly sketch how that is done.

As documented by the client-server based STUN[45] protocol, there are four types of NAT in common use today. Of these four types, all have the property that if a UDP packet is sent from IP address A port $p_a$ to IP address B port $p_b$, the NAT device will allow packets from IP address B port $p_b$ to flow to IP address A port $p_a$. Any system that does not permit this is broken because it is allowing outbound packets without allowing any response to those packets, which meets the requirements of almost no applications. In addition to the above property, three out of four of the common NAT types (all but the symmetric) use the same mapping for the NAT's port $\rightarrow$ internal *(IP, port)* pair. Thus each connection in the P2P network is an opportunity for a node to discover if any IP translation is going on, and if so record its translated address. Once a node discovers that its address is being translated, it advertises that translated address to another nodes over the Brunet network. Finally, since the Brunet connection protocol specifies that each node try to contact the other, to the firewall, one of the packets will appear to be the response to a previous request and thus will be allowed to pass.

This NAT traversal protocol makes use of the same facts about NATs that the STUN[45] protocol uses. Furthermore, this approach is decentralized and introduces no single points of failure or dedicated servers (unlike the STUN protocol).

While this NAT/firewall traversal may seem like a subversion of network policies, this is arguably not the case. In fact, both nodes in this scenario are actively attempting to contact the other, and as such, any unrequested packets are still filtered by the NAT or firewall.

### E. Multiple IPs and mobility

The current solution of mapping IP addresses to Brunet addresses using SHA-1 hashes requires one P2P node per IP address. Because of this requirement, a single IPOP node running on a host cannot "route for" multiple virtual IPs (eg. multiple VMs hosted by a physical machine). The problem is aggravated when virtual IP addresses are mobile — a situation that can occur when virtual machines are allowed to migrate ([47], [13]). A solution to this problem involves using the P2P system as a distributed hash table (DHT) [52][54][42].

We call our proposed protocol for mapping an IP destination to a Brunet address "Brunet-ARP". An IPOP node informs about each virtual IP address it "routes for" to the Brunet node whose address is 160 bit SHA-1 hash of that IP address. We call this node "Brunet-ARP-Mapper". Now when a node has an IP packet to send, it inquires about the brunet destination from the corresponding "Brunet-ARP-Mapper" whose Brunet address is the 160 bit SHA-1 hash of the destination IP address. This information can then be cached at the source node. When a VM migrates (retaining its IP address), the information is updated at its corresponding "Brunet-ARP-Mapper".

The issues related to the "Brunet-ARP-Mapper" for an IP address being down have been dealt with in the DHT literature [52][54][42]; and IPOP too can benefit from these solutions.

### IV. EXPERIMENTS

In this section we present and discuss a series of experiments that have been used to evaluate the performance of the current IPOP prototype. The purpose of these experiments are to show the feasibility of overlaying IP over P2P using current technologies, and to highlight the applications that can benefit from the IPOP architecture. It is worth noting that the P2P network *has not been optimized* in any form to support IP traffic.

To evaluate the performance of our system, we have conducted a variety of experiments. First, we evaluate latency and throughput of a single hop of the overlay network (Section IV-B). We also report on the performance of multi-hop routing on a larger-scale overlay deployed on top of the Planet-Lab [7] testbed (Section IV-D). These experiments quantify the overhead of the user-level routing layer. We also performed experiments to evaluate the performance of an MPI-based parallel application running on nodes interconnected by IPOP (Section IV-C).

### A. Experimental setup

Figure 4 shows the experimental set up we used for our measurements in both LAN and WAN environments. The LAN testbed consist of F1, F2, F4 that reside in the ACIS laboratory private network at the University of Florida. The machine F1 is a VM based on VMware GSX Server 2.5.1 (build 5336), running on a dual (hyperthreaded) Intel Xeon 2.40 GHz host. The machine F2 is a physical host with Intel Pentium III 1122 MHz processor. The machine F4 is a VM based on VMware ESX Server 2.1.1 (build-9157) running on Intel Xeon 2.00 GHz host. It has a private interface (connected to the LAN) and a public interface (connected to the campus public network). Machine F3 is in a different University of Florida LAN, and is a VM based on VMware GSX Server 3.1.0 (build-9089), running on a dual (hyperthreaded) Intel Xeon 3.20 GHz host. In the WAN testbed, we have two machines V1 and L1 which are situated at Virginia Institute of Marine Sciences and Louisiana State University, respectively. Both V1 and L1 machines are behind site firewalls. These machines are connected to the machines at University of Florida via Abilene. V1 is a is a 4-way (hyperthreaded) Intel Xeon 2.8 GHz host. L1 is a VM based on VMware GSX Server 3.1.0 (build-9089), running on a dual (hyperthreaded) Intel Xeon 3.20 GHz host. All the machines run Linux (kernel 2.4).

### B. Analysis: Link latency and throughput

We measured latency by using the round-trip delay of an ICMP request/response pair. We computed the average and standard deviation of 1000 measurements. For the LAN case, we measured the round-trip delay between machines F2 and F4 (Figure 4), while for the WAN case, we measured the round-trip delay between machines F4 and V1. Experiments were performed with IPOP routing over both TCP and UDP, leveraging the support for both protocols in Brunet. It should be noted that IPOP-TCP and IPOP-UDP experiments were performed at different times; due to differences in host load (which influences IPOP overhead at P2P routers and endpoints) and network conditions, different set of values were observed for the physical network, both of which have been reported in this paper.

Table I summarizes the results for the LAN and WAN latency experiments. In this table, the experiment is setup with Brunet nodes connected by TCP. Table I summarizes the results of an experiment with Brunet nodes connected by UDP. We observe the latency to be of the order of 6-10ms per ICMP packet when using IPOP. Latencies of the order of milliseconds/packet have also been reported in context of other user-level routing systems, such as VNET [13]. The LAN experiment provides a rough estimate of the overhead associated with our implementation of IPOP. We attribute this overhead to the traversal of kernel TCP/IP stack twice by any packet sent on the virtual network (once on the virtual interface, and additionally on the physical interface). While the relative overhead is high in the LAN environment, for the WAN used in this experiment the overhead is 33% of that of the physical network for both TCP and UDP implementations of IPOP. In a WAN, the overhead of user-level routing gets amortized over the number of physical hops (in our case, 10) that make up a P2P link.

We have measured the average throughput of an overlay link using the ttcp program, which is commonly used to test TCP performance in IP networks. It times the transmission and reception of data between two systems. For a LAN, we measured the throughput between nodes F2 and F4 (Figure 4), while for WAN case, we measured the throughput between machines F4 and V1. In Table II, we compare the throughput of a single overlay link (TCP and UDP) over a LAN to that of the physical network, and we observe it to be only 20% of that of the physical network. In Table III, we measure the throughput for with two transfer sizes in a WAN, and interestingly the overlay link could harness up to 80% of the capacity of the physical network. Such observations on higher UDP throughput over TCP have also been reported in context of VNET in [36]. The measuremts were conducted on the physical machines (F2 for LAN and V1 for WAN), but the other side of the connection (F4) was a VM based on VMware ESX 2.1.1 with 100 Mbps virtual ethernet card.

### C. Analysis: MPI application

We conducted an experiment with an MPI-based application called LSS (Light Scattering Spectroscopy) [41]. LSS is an application in biomedical engineering that analyzes of a set of spectral images

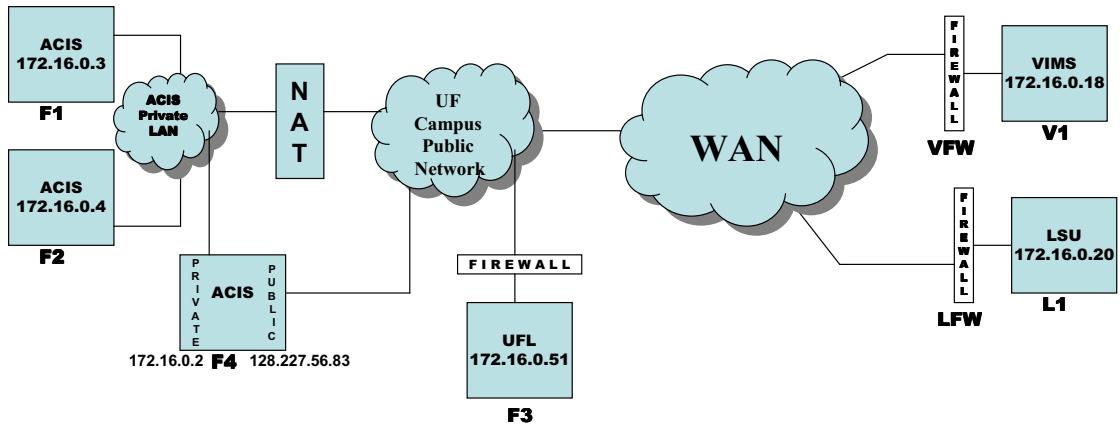

Fig. 4. Experimental Setup. The testbed consists of six machines (F1, F2, F3, F4, V1, L1) distributed across three sites. For TCP, firewalls VFW and LFW do not allow incoming connections from machines F1, F2, F4, and only allow incoming connections from machine F3 for port 22 (SSH). VFW allows outgoing connections to machines F1..F4; however, LFW only allows outgoing TCP connections to machine F3.

|     |          | mean (msec) | std. dev (msec) |          | mean (msec) | std. dev (msec) |
|-----|----------|-------------|-----------------|----------|-------------|-----------------|
| LAN | physical | 0.898       | 2.843           | physical | 0.625       | 0.214           |
|     | IPOP-TCP | 7.832       | 21.704          | IPOP-UDP | 6.859       | 3.180           |
| WAN | physical | 38.801      | 6.541           | physical | 34.492      | 0.702           |
|     | IPOP-TCP | 48.539      | 3.117           | IPOP-UDP | 45.896      | 9.782           |

TABLE I

MEAN AND STANDARD DEVIATION OF 1000 PING ROUND-TRIP TIMES OF TCP AND UDP VERSIONS OF IPOP WITH PHYSICAL NETWORK.

|          | Abs. b/w (KBps) | Rel. b/w (IPOP/Phys) |
|----------|-----------------|----------------------|
| physical | 8255            |                      |
| IPOP-TCP | 2389            | 29%                  |
| physical | 9416            |                      |
| IPOP-UDP | 1905            | 20%                  |

TABLE II

COMPARISON OF THROUGHPUT OF A SINGLE OVERLAY LINK (TCP AND UDP) IN LAN TO THAT OF PHYSICAL NETWORK, MEASURED USING TTCP; TRANSFER SIZE = 92.97 MB

|                | Abs. b/w (KBps) | Abs. b/w (KBps) | Rel. b/w IPOP/Phys | Rel. b/w IPOP/Phys |
|----------------|-----------------|-----------------|--------------------|--------------------|
| File size (MB) | 13.09           | 92.97           | 13.09              | 92.97              |
| physical       | 1419            | 1419            |                    |                    |
| IPOP-TCP       | 673             | 688             | 47%                | 48%                |
| physical       | 1538            | 1531            |                    |                    |
| IPOP-UDP       | 1239            | 1150            | 81%                | 75%                |

TABLE III

COMPARISON OF THROUGHPUT (IN KBPS) OF A SINGLE OVERLAY LINK (TCP AND UDP) OVER WAN TO THAT OF A PHYSICAL NETWORK FOR TRANSFER SIZES OF 13.09 MB AND 92.97 MB

obtained experimentally from a tissue sample against a set of database files containing known spectra generated analytically using Mie scattering theory. The application finds the analytical spectrum that best fits the experimental data by applying a least-square fitting algorithm for each database record and selecting the fit with minimum error across all records. It is an application that exhibits both compute-intensive behavior (in the computation of the least-square fits) and data-intensive behavior (in the access of large lookup database files).

LSS has been parallelized by distributing the least-square computation across multiple processors using message-passing (MPI) libraries. The parallel version of LSS involves a master node and one or mode slave nodes. The analysis of each image against different databases takes place in parallel, local results are communicated back to the master which uses these results to compute the image parameters. SSH is required to start the lam daemons on each compute node before parallel execution begins. The goal of this experiment was to demonstrate our ability to run an application requiring MPI, SSH, and NFS over nodes connected through IPOP. Without network virtualization provided by IPOP, it would not have been possible to run parallel-LSS because of inadequate node connectivity of the

| # of nodes | Image 1 | Images 2-6 | Total |
|---|---|---|---|
| 1 | 811s | 834s | 1645s |
| 4 | 378s | 217s | 595s |

TABLE IV

EXECUTION TIMES (IN SECONDS) FOR LSS IMAGE ANALYSIS FOR BOTH SEQUENTIAL (1 NODE) AND PARALLEL (4 NODES) EXECUTIONS. EXECUTION TIMES ARE COLLECTED FOR SIX CONSECUTIVE LSS ANALYSIS RUNS USING SIX DIFFERENT IMAGES. THE TIME TO PROCESS THE FIRST IMAGE (WHEN NFS CACHES ARE COLD) IS REPORTED SEPARATELY.

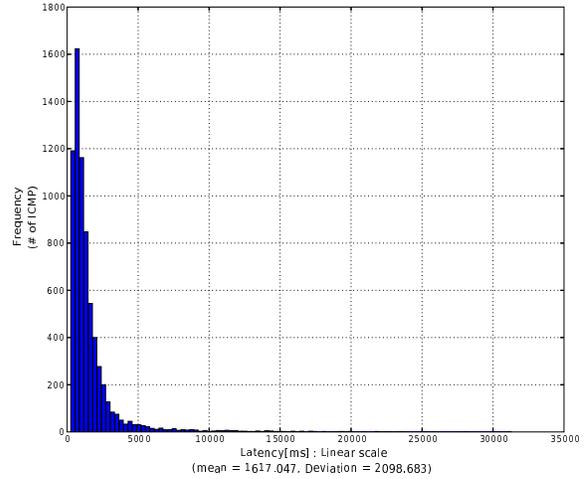

Fig. 5. Distribution of round-trip latencies for ICMP/ping packets over 118-node Planet-Lab overlay. Two hops separate the ping source from the destination.

physical network.

In this experiment, the machine F4 in Figure 4 is used as a central file server where the image and database files, LSS and MPI (LAM 7.0.6 implementation) binaries are stored. This data is available to the other compute servers through an NFS-based virtual file system [40], which provides transparent user-level client-side disk caching that exploits the temporal locality of references available across runs of LSS that analyze different images using the same databases.

In the experiment, we run the LSS analysis for six different images and four database files each of size 32 MB. Table IV shows the LSS execution times for both sequential (1 node) and parallel executions (4 nodes). The analysis of first image is slow because initially the client-side NFS caches are cold, and nodes spend most of the time doing remote I/O to fetch corresponding database files. For subsequent images (2 through 6), the database files are available in the local cache at each compute node, and the parallel execution achieves a speed up of 3.8 over the sequential execution.

### D. Analysis: Multi-hop routing

We also conducted latency experiments on a TCP-based Brunet P2P network consisting of 118 nodes established across Planet-Lab. The experiment was setup by first deploying the nodes over Planet-Lab. The IPOP routing layer is self-configured by Brunet as each node joins/leaves. We then connected two of our testbed nodes (F2, F4) to this overlay network and measured the round-trip ping times between them. Figure 5 shows the distribution of 10000 ping messages collected in one experiment where there were two overlay hops between the the source and destination.

An interesting observation was that forward and backward paths between F2 and F4 were different. The ICMP request messages (from F2 to F4) went through the Planet-Lab node named planetlab15.millenium.berkeley.edu, while the responses came through pli1-br-2.hpl.hp.com.

The results show average ping times in excess of 1.6 seconds. From pings to the machines (planetlab15.millenium.berkeley.edu, pli1-br-2.hpl.hp.com) over the physical network, we observed that the average IPOP overhead was approximately 1.4 seconds over that of the same path on the physical network. We attributed this overhead to the high load under which Planet-Lab nodes were subject to at the time of the experiment. Monitoring of the intermediate Planet-Lab nodes that were used by IPOP to route ICMP packets in this experiment shows that the CPU load was in the excess of 10. Because IPOP runs at user-level and competes for CPU time with other Planet-Lab tasks in a heavily-loaded environment, its performance suffers. Our goal in this experiment was to show the feasibility of applying IPOP to a large distributed network, and not to characterize IPOP's performance over Planet-Lab. Nonetheless, the Planet-Lab experiments provide directions for future work towards improvements in performance for IPOP that are driven by IP routing needs.

### V. DISCUSSION

The Planet-Lab experiments showed that, although IP-level connectivity was established by IPOP, the performance delivered by the system suffered because of (a) contention for resources at intermediate routing nodes, with CPU loads in excess of 1000%, and (b) long round-trip latencies (in excess of 100ms) in the physical network among nodes that were used for routing. Currently, IPOP does not account for host and link performance for routing, hence choices of intermediate nodes that yield poor performance for IP routing can be made.

These results motivate research in P2P algorithms that account for *physical network* performance data (link latency and bandwidth, node load) to establish routing paths that adapt dynamically to network conditions.

We consider the following extensions to improve P2P routing performance:

*1) Short-cut connections:* We can extend Brunet to support monitoring of P2P traffic at each overlay node, and provide for setting up direct edges (if possible) when communication between a pair of nodes exceeds certain threshold. This is equivalent to using IP routing between those two overlay nodes while the P2P provides for address resolution and boot-strapping of such short-cut connections. Such enhancement to Chord [52] lookup protocol has been implemented in i3 [51].

*2) Single TCP/IP protocol stack traversal:* The high overhead incurred on LAN is mainly because each packet has to traverse the kernel TCP/IP stack twice (once on the virtual interface, and additionally on the physical interface). Typical Grid computing systems are based on clusters, in many cases featuring high-performance network cards [1][5] that support a user-level communication architecture [15] that avoids the overhead of traversing the kernel TCP/IP stack. IPOP nodes can be enhanced to discover if user-level communication libraries are available, and take advantage of the functionality provided by such cards to bypass one kernel TCP/IP stack. Through virtualization, applications running on IPOP overlays would be oblivious to the choice of tunneling over TCP/IP on a WAN or over a user-level communication architecture on a cluster LAN.

## VI. CONCLUSIONS

In this paper, we have described a novel network virtualization technique - IPOP - which allows aggregating resources spanning multiple domains (even behind firewalls, NATs) into a single virtual network, through the use of virtual devices and P2P networks. IPOP preserves the TCP/IP protocol stack semantics; this feature, coupled with the bidirectional connectivity it provides, enables unmodified distributed applications (written for LANs) to run seamlessly on WANs, over the virtual network. IPOP leverages the self-configuring, scalable and fault-tolerant nature of P2P networks to achieve overlay routing without centralized administrative control.

We have also evaluated the overheads associated with our current prototype to establish the feasibility of this approach. Experimental results show that (1) the average latency overheads for a single hop are in the range of 6-10ms, which is acceptable for many WAN applications; (2) the average ttcp throughput delivered by IPOP in a WAN scenario is as high as 80% of the physical network's bandwidth (over Brunet-UDP); (3) IPOP successfully provided a self-configured overlay that efficiently supported the execution (across firewalled nodes) of a parallel application that uses several TCP/IP services, and (4) IPOP successfully self-configured a 118-node overlay that supported 2-hop virtual IP routing over the Planet-Lab testbed.

## VII. ACKNOWLEDGMENTS

Effort sponsored by the NSF under grants EIA-0224442, EEC-0228390, ACI-0219925, ANI-0301108 and SCI-0438246 and carried out as a component of the "SURA Coastal Ocean Observing and Prediction (SCOOP) Program", an initiative of the Southeastern Universities Research Association (SURA). Funding support for SCOOP has been provided by the Office of Naval Research, Award# N00014-04-1-0721 and by the NOAA Ocean Service, Award # NA04NOS4730254. The authors also acknowledge a SUR grant from IBM. Any opinions, findings and conclusions or recommendations expressed in this material are those of the authors and do not necessarily reflect the views of the sponsors. The authors would like to thank Justin Davis, Vladimir Paramygin, Chirag Dekate and David Forrest for assisting in the configuration of resources.